\documentclass[11pt]{cernrep}
\usepackage{epsfig}
\usepackage{graphicx}
\usepackage{here}
\begin{document}
\pagestyle{plain}

\title{LEADING ORDER PQCD HADRON PRODUCTION AND NUCLEAR  \\[.5ex] 
       MODIFICATION FACTORS AT RHIC AND THE LHC~\footnote{ \ \ Contribution to 
the CERN  Yellow Report on Hard Probes in Heavy Ion Collisions at the LHC. 
This manuscript also provides details on the background calculations for 
Ref.~\cite{Vitev:2002pf} and discussion on results that have been omitted
there due to space limitations.}}

\author{Ivan~Vitev$^{\, \dagger}$}

\maketitle

\hspace*{-.6in} \begin{tabular}{rl}
\noindent ${^\dagger \!\!\!\!\!}$&Department of Physics and Astronomy \\
                          & Iowa State University, Ames, IA 50011 \\
                          & E-mail: ivitev@iastate.edu
\end{tabular}

\noindent \maketitle

\begin{abstract}
Hadron production in leading order pQCD is reviewed. 
The shape of the single inclusive particle spectra is well 
described for $p_T \geq 2-3$~GeV at  center of mass energies 
from $20$~GeV to  $2$~TeV.  The phenomenological  
K-factor is found to decrease systematically with $\sqrt{s}$. 
For ultra-relativistic heavy ion reactions
the calculation is augmented with the effects of initial multiple
parton scattering and final state radiative energy loss.    
Baseline CERN-LHC predictions for hadron production in $p+p$ 
and suppression in central $Pb+Pb$ reactions at 
$\sqrt{s} = 5.5$~TeV are given in comparison to the  
corresponding results at BNL-RHIC  and CERN-SPS energies. 
  
\end{abstract}

\section{INTRODUCTION}

One of  the main  goals of the upcoming $p+p$ program at $\sqrt{s} = 14$~TeV
at the Large Hadron Collider  (LHC) at CERN is the unambiguous 
discovery of  the Higgs boson, predicted by the standard model of 
particle interactions, as well as the search for physics that reaches 
beyond our current understanding of the constituents of matter 
and the force mediators. Equally  important, however, is the continuing effort 
to investigate the strong sector of the SM and probe experimentally 
some of the  fundamental predictions of QCD: the deconfinement phase 
transition and chiral symmetry restoration.
To  order   $(\alpha^{LO}_s)^2$ the strong coupling constant   
$\alpha_s=g_s^2/4\pi$ reads 
\begin{eqnarray}
\alpha_s (Q^2,n_f) = \alpha^{LO}_s (Q^2,n_f) \left[  
1-\frac{1}{4\pi} \frac{ 102-\frac {38}{3} n_f }{ 11-\frac {2}{3} n_f }
\alpha^{LO}_s (Q^2,n_f) \ln \, \ln  \frac{Q^2}{\Lambda^2_{QCD}} 
\right] \;, \quad
\label{alpha-lo2}
\end{eqnarray}
where the lowest order $\alpha^{LO}_s (Q^2,n_f) =  {4\pi} /
{\left ( 11-\frac{2}{3} n_f \right )
\ln  \frac {Q^2}{\Lambda^2_{QCD}}  }.$
QCD is thus  ``asymptotically free'', i.e.  
for $n_f \leq  6$ the running coupling, Eq.~(\ref{alpha-lo2}), 
approaches zero in  the limit of large momentum transfer $Q$, - 
a general  feature of non-Abelian 
gauge theories with sufficiently small number of fermions.  
In hot and dense matter the typical momentum scale is on the order 
of the temperature $Q \sim T$, assuming local thermal equilibrium. 
Large initial temperatures and energy densities  can be  experimentally 
achieved in ultra-relativistic heavy ion reactions.  
The $Pb+Pb$  program at $\sqrt{s}_{NN} = 5.5$~TeV at the LHC is targeted 
at the search for the quark-gluon plasma (QGP)~\cite{Collins:1974ky}   
and the study of its properties. 
The Bjorken-estimated  $T_i \simeq 1$~GeV for these conditions  
exceeds by a large margin the current lattice QCD results for the critical 
temperature $T_c \simeq 170$~MeV. An important advantage of the LHC is that 
it will ultimately make small-$x$ physics studies in heavy ion collisions 
feasible (not applicable at RHIC except possibly for $p_T < 0.5$~GeV). 
At very small values of $x$ the rapid growth  of the gluon  distribution 
in nucleons  and nuclei  is tamed by absorption terms that lead to a 
modification~\cite{Mueller:wy} of the DGLAP evolution equations and
correct the small-$x$ unitarity problem:     
\begin{eqnarray} 
\frac{\partial xG(x,Q^2)}{\partial \ln Q^2} 
&=& \frac{C_A \alpha_s}{\pi}
\int_x^1 \frac{dx^\prime}{x^\prime}\frac{x}{x^\prime}
\gamma^{gg}\left( \frac{x}{x^\prime} \right)
x^\prime G(x^\prime,Q^2)  \nonumber \\ 
&& -  \frac{4 \pi^3}{N_c^2-1} 
\left(\frac{C_A \alpha_s}{\pi} \right)^2
\frac{1}{Q^2} \int_x^1 \frac{dx^\prime}{x^\prime} (x^\prime)^2 
 G^{(2)}(x^\prime,Q^2) \;, 
\label{mq}
\end{eqnarray}
where $\gamma^{gg}$ is the gluon splitting function and $G^{(2)}(x,Q^2)$ 
is proportional to the gluon density overlap.
An opportunity to test these predictions  at the LHC is provided by
the $p+Pb$ program at $\sqrt{s}=8.8$~TeV since it stands the best 
chance of  identifying initial state nuclear 
effects~\cite{Dumitru:2002qt} and  separating 
them~\cite{Vitev:2002pf} from the final state multi-parton 
interactions.

The search of novel physical effects in ultra-relativistic heavy ion 
reactions at the LHC can only rely on a detailed comparison 
between the experimental data and the projected current, or ``conventional'', 
knowledge. This calls for detailed baseline calculations of jet and hadron 
production at those center of mass energies as well as an  estimate 
of the known nuclear effects. The purpose of this manuscript 
is to present a {\em lowest order} (LO) 
analysis of inclusive hadron production   up to the Tevatron 
energies and discuss hadron differential cross sections
and composition  at the LHC. This choice is dictated by the requirement
of self-consistent incorporation of nuclear effects that are at present 
computed/parameterized to LO. It also complements next-to-leading 
order (NLO)  calculations of jet and hadron 
production~\cite{Accardi:2002vt,nloh}. Evaluation of the nuclear 
modification factors at the LHC in comparison to RHIC  
and discussion of the hadron composition is also presented.
Results on Cronin and shadowing effects are given at forward 
(in the direction of the proton/deuteron beam) $y=+3$ rapidity.

\section{HADRON PRODUCTION IN FACTORIZED PQCD}

The standard  factorized pQCD hadron production formalism 
expresses the differential 
hadron cross  section in $N+N \rightarrow h+X$  as a convolution of 
the measured parton distribution functions (PDFs)  
$f_{\alpha/N}(x_\alpha,Q_\alpha^2)$  for the interacting 
partons ($\alpha = a,b$), 
with the fragmentation function  (FFs) $D_{h/c}(z,Q^2_c)$ for
the leading scattered parton  $c$  into a hadron of flavor $h$ and the
parton-parton differential cross sections for the elementary sub-process 
$d\sigma^{(ab \rightarrow cd)}/d\hat{t}$: 
\begin{eqnarray} 
E_{h}\frac{d\sigma^{NN}}{d^3p} &=&
K_{NLO}   \sum_{abcd}\,  \int\limits_0^1  dz_c  
\int\limits_{x_{a  \min}}^1 \int\limits_{x_{b  \min}}^1dx_a  dx_b \;
f_{a/p}(x_a,Q^2_a) f_{b/p}(x_b,Q^2_b) \nonumber \\
&& \times \; D_{h/c}(z_c,{Q}_c^2) 
 \frac{\hat{s}}{\pi z^2_c} \frac{d\sigma^{(ab\rightarrow cd)}}
{d{\hat t}} \delta(\hat{s}+\hat{u}+\hat{t}) \; .
\label{hcrossec}
\end{eqnarray}
A list of the lowest order partonic cross sections can be found 
in~\cite{Owens:1986mp}. In Eq.~(\ref{hcrossec}) $x_a, x_b$ are the
initial  momentum  fractions  carried  by the interacting partons 
and  $z_c=p_h/p_c$  is  the momentum fraction of the observed hadron. 
$K_{NLO}$ is a phenomenological factor that is meant to account 
for next-to-leading order (NLO) corrections. It is $\sqrt{s}$ 
and scale dependent  and takes typical values $ \simeq 1-4$.
One usually finds that Eq.~(\ref{hcrossec}) over-predicts 
the  curvature  of the inclusive hadron spectra 
$ | \partial_{p_T} d\sigma^h |$  at transverse momenta $p_T \leq 4$~GeV. 
This can be partly corrected by the introduction of a small 
intrinsic (or primordial) $k_T$-smearing of partons, transversely 
to  the collision axis, and generalized parton distributions
$\tilde{f}_\alpha(x,k_T,Q^2)$  motivated by the pQCD initial state 
radiation. For the corresponding modification of the kinematics
in (\ref{hcrossec}) in addition to the $\int d^2 k_T^a 
\int  d^2 k_T^b \,(\cdots) $ integrations  see~\cite{Owens:1986mp}.
The generalized parton distributions are often approximated as
\begin{equation}
\tilde{f}_\alpha(x,k_T,Q^2) \approx f_\alpha(x,Q^2) g(k_T), \quad 
g(k_T) = \frac{e^{-k_T^2/\langle {k}_T^2 \rangle}}
{\pi\langle {k}_T^2 \rangle } \; ,
\label{gassm}
\end{equation}
where the width  $\langle {k}_T^2 \rangle$ of the Gaussian enters as a 
phenomenological parameter.

\vskip 0.5cm
\begin{figure}[htb!]
\begin{center} 
\hspace*{-0.2in} 
\epsfig{file=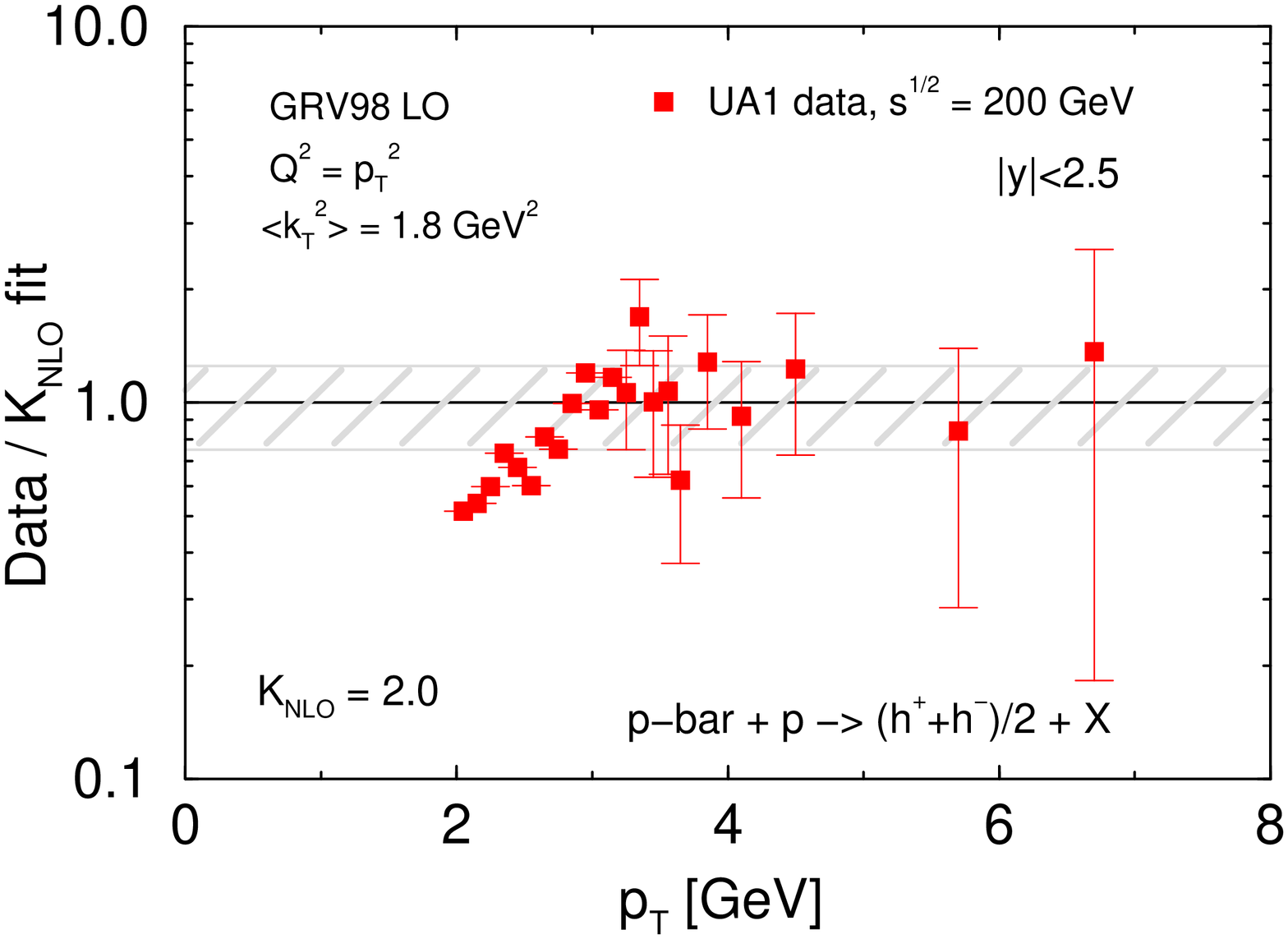,height=3.2in,width=2.6in,
bbllx=90,bblly=0,bburx=600,bbury=700, clip=,angle=-90}
 \epsfig{file=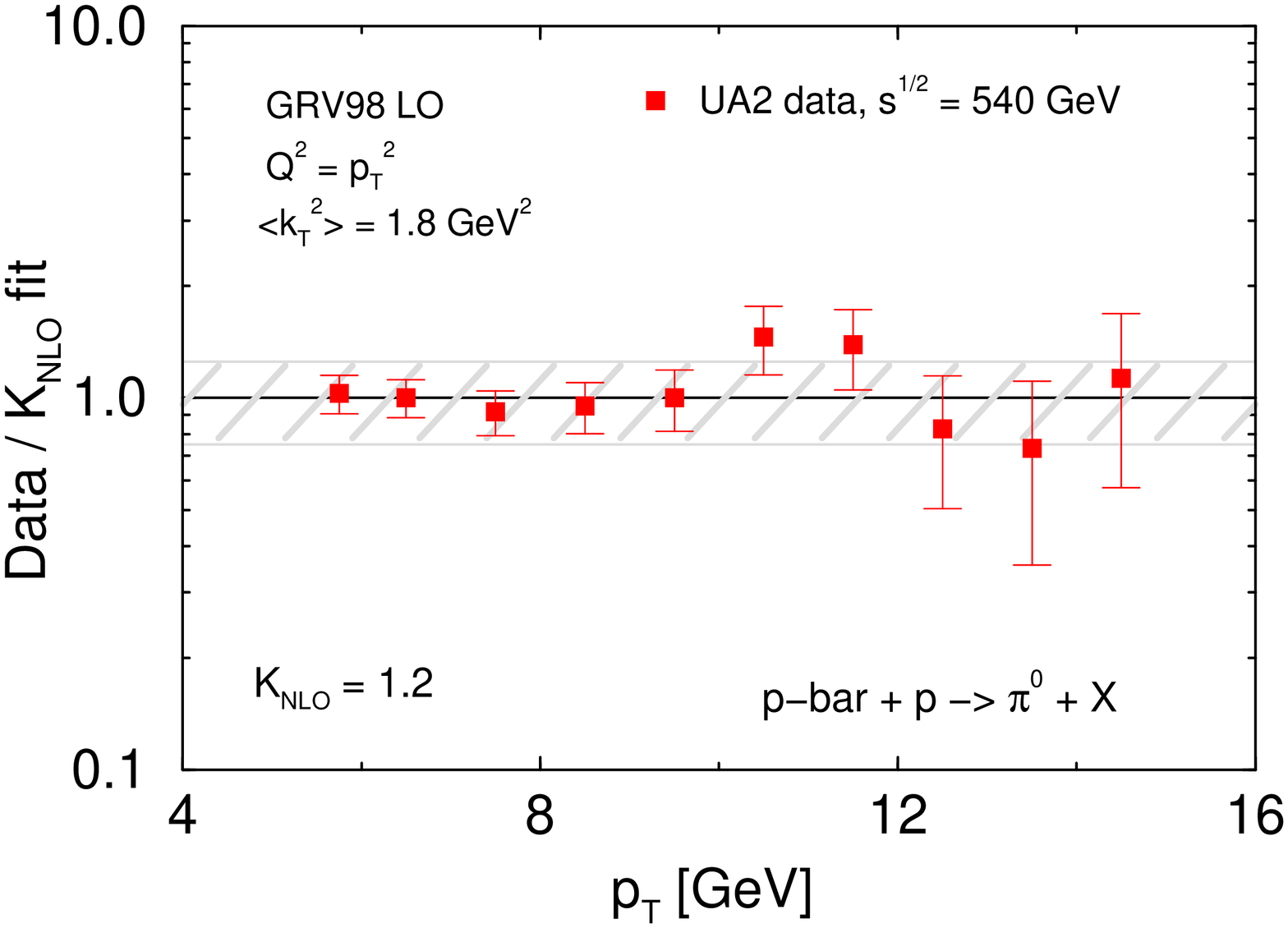,height=3.2in,width=2.6in,
bbllx=90,bblly=0,bburx=600,bbury=700, clip=,angle=-90}
\hspace*{-0.2in} 
\epsfig{file=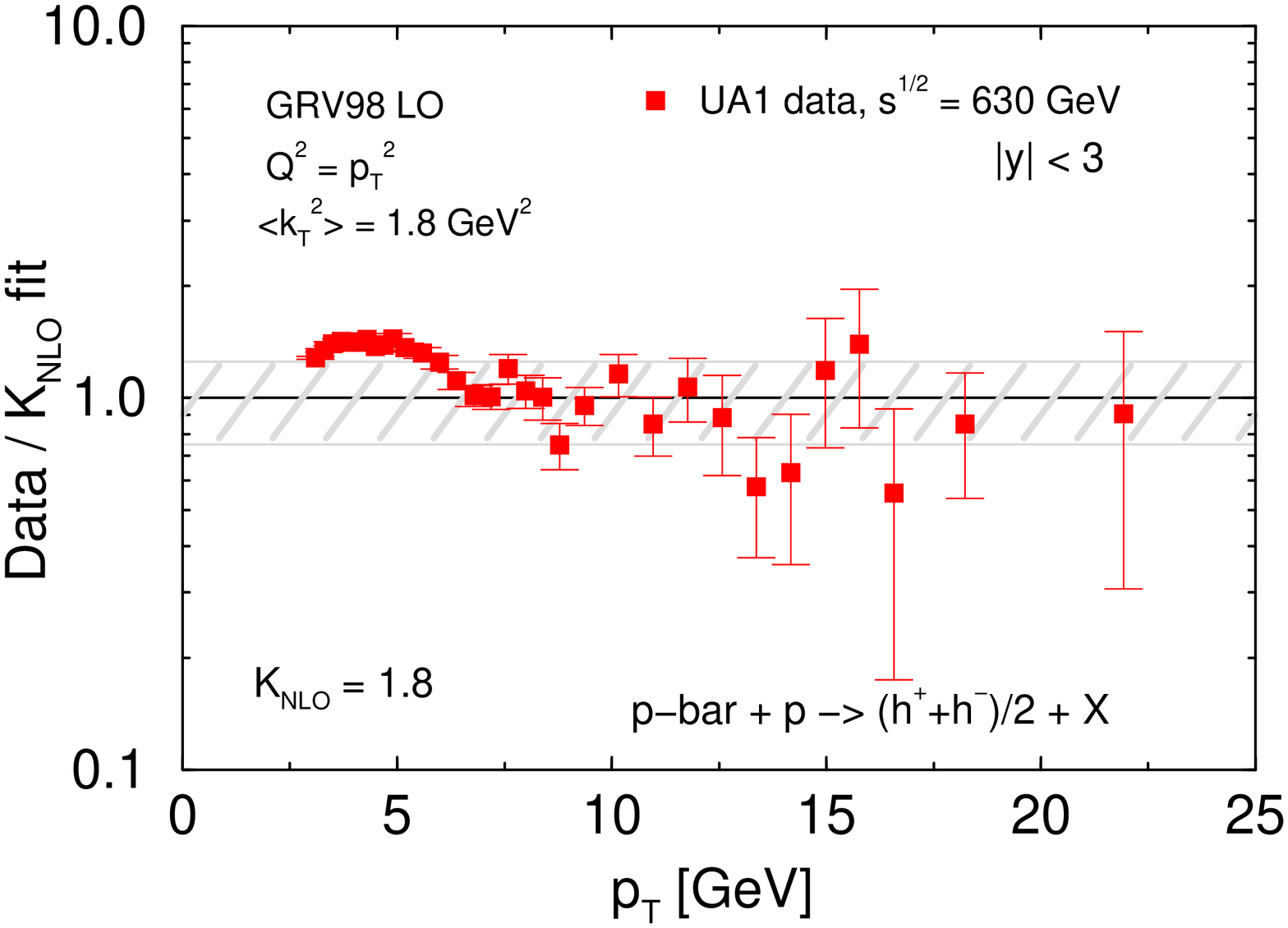,height=3.2in,width=2.6in,
bbllx=90,bblly=0,bburx=600,bbury=700, clip=,angle=-90}
 \epsfig{file=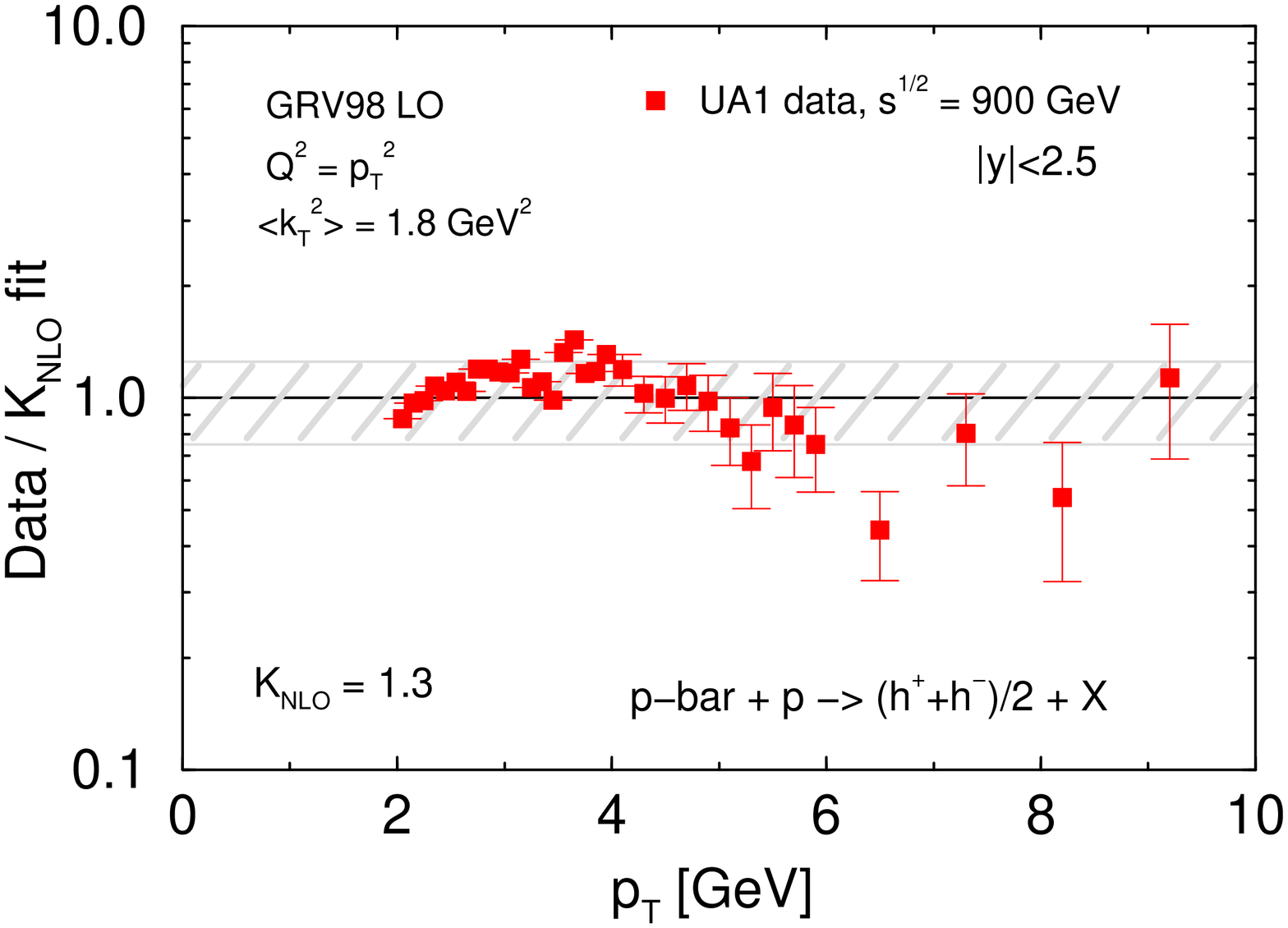,height=3.2in,width=2.6in,
bbllx=90,bblly=0,bburx=600,bbury=700, clip=,angle=-90}
\hspace*{-0.2in} 
\epsfig{file=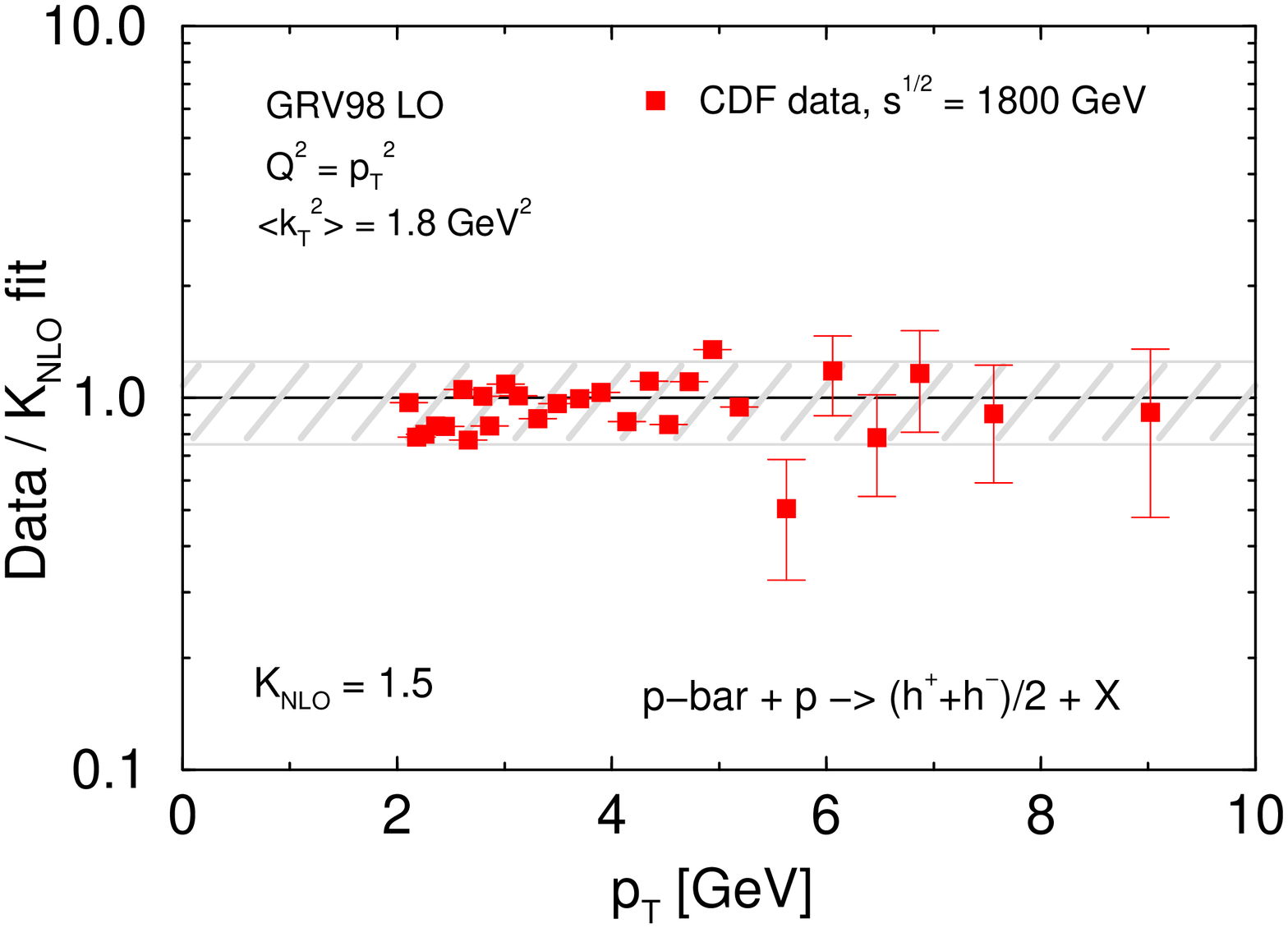,height=3.2in,width=2.6in,
bbllx=90,bblly=0,bburx=600,bbury=700, clip=,angle=-90}
 \epsfig{file=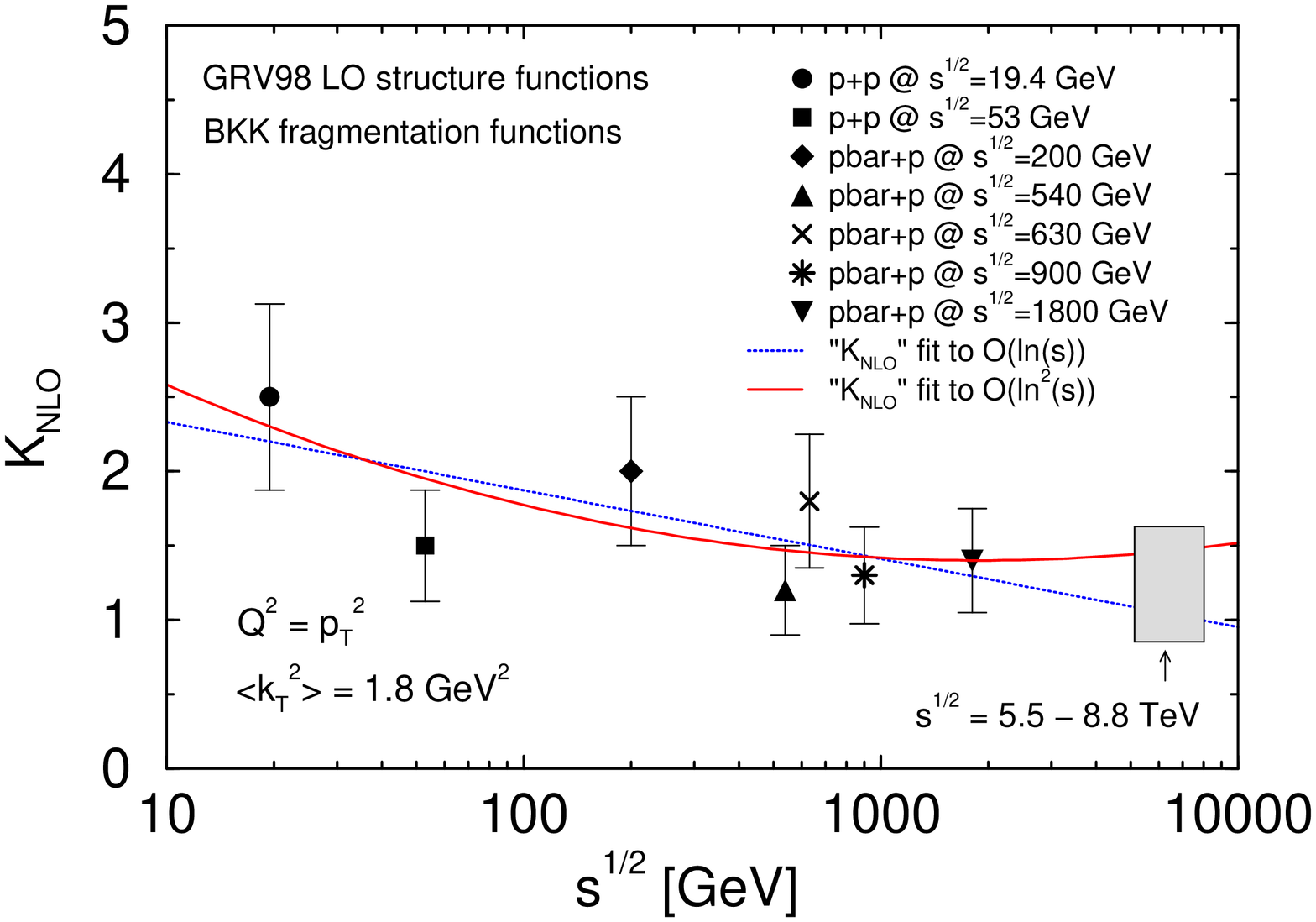,height=3.2in,width=2.6in,
bbllx=90,bblly=0,bburx=600,bbury=700, clip=,angle=-90}
\vspace*{-0.1in}
\caption{\small  Extracted  $K_{NLO}$  from comparison of LO 
pQCD calculation to data~\protect{\cite{Antreasyan:cw,Alper:nv,Albajar:1989an,Banner:1984wh,Bocquet:1995jr,Abe:yu}  } at and about mid-rapidity  in the 
range $2\leq p_T \leq 25$~GeV. A systematic decrease of $K_{NLO}$ 
with $\sqrt{s}$ is observed and illustrated in the bottom right panel. The 
projected 50\% uncertainty at $\sqrt{s}= 5.5 -8.8$~TeV is also shown. }     
\label{lhc-h:fig1}
\end{center} 
\end{figure}
Perturbative QCD fits to data~\cite{Cronin:zm,Straub:xd,Antreasyan:cw,Alper:nv,Albajar:1989an,Banner:1984wh,Bocquet:1995jr,Abe:yu}  
use  different coupled choices for  $K_{NLO}$ and 
$\langle {k}_T^2 \rangle$ and  the extracted values are thus not 
directly comparable. However, similar agreement 
between data and theory at the level of spectral shapes and  the $\sqrt{s}$ 
dependence of the corrective factors discussed above is found.
In~\cite{Zhang:2001ce} the factorization and fragmentation scales 
were set to $Q_{PDF}=p_T/2$  and  $Q_{FF}=p_T/2z_c$ and no $K_{NLO}$ 
factors were employed.  The extracted  $\langle {k}_T^2 \rangle$  
decreases from $2.7$~GeV$^2$  at $\sqrt{s}\simeq 50$~GeV to  
$0.75$~GeV$^2$ at  $\sqrt{s}\simeq 2$~TeV. Alternatively, 
in~\cite{Eskola:2002kv} no primordial $k_T$-smearing was used and the 
scales in the calculation were fixed to be  $Q_{PDF}=Q_{FF}=p_T$. The 
deduced  $K_{NLO}$ decreases from  $\sim 6$ at $\sqrt{s}\simeq 50$~GeV 
to $\sim 1.5$ at   $\sqrt{s}\simeq 2$~TeV.

In the fits shown in Fig.~\ref{lhc-h:fig1} we have used  the GRV98 LO 
PDFs~\cite{Gluck:1998xa} and the BKK LO FFs~\cite{Binnewies:1994ju}. 
Proton+antiproton fragmentation has been  parameterized  as 
in~\cite{Wang:1998bh}, inspired from PYTHIA~\cite{Sjostrand:1993yb}   
results.  A fixed $\langle {k}_T^2 \rangle_{pp}=1.8$~GeV$^2$ has been 
employed, leading to  a $K_{NLO}$ parameter  that  naturally exhibits a 
smaller variation with $\sqrt{s}$. A $\pm 25 \%$ error band 
about the $K_{NLO}$ value, fixed by the requirement to match the moderate- 
and high-$p_T$  behavior of the data, is also shown. The  
fragmentation and factorization scales were fixed as 
in~\cite{Eskola:2002kv}. In the lower right  panel the systematic 
decrease of the next-to-leading order K-factor is
presented. Two fits to $K_{NLO}$ have been used: linear  
$K_{NLO} = 2.7924-0.0999 \ln s $ and quadratic   
$K_{NLO} = 3.8444-0.3234 \ln s + 0.0107 \ln^2 s$ in  $\ln s$.  
For center of mass energies up to 1~TeV the two parameterization  differ
by less than 15\% but this difference is seen to grow to 30\%-50\% at 
$\sqrt{s}=5-10$~TeV.

\vskip 0.5cm
\begin{figure}[htb!]
\begin{center} 
\hspace*{-0.2in}\psfig{file=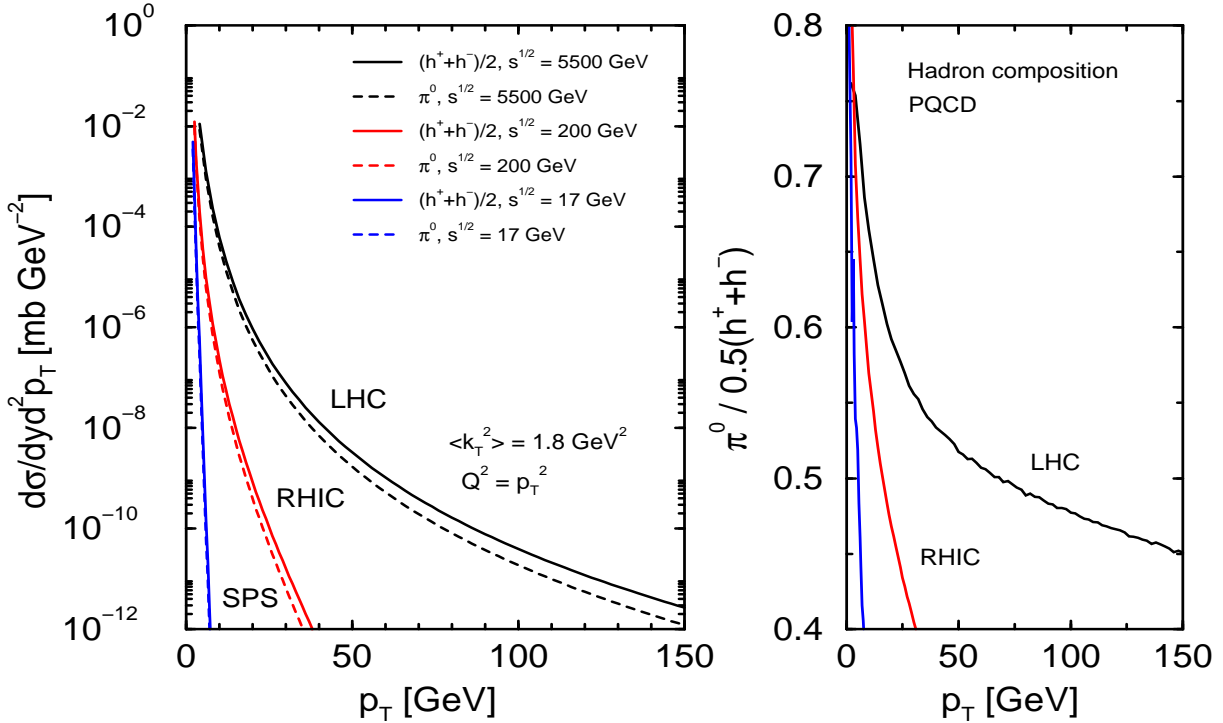,height=6.9in,width=4.in,
bbllx=90,bblly=0,bburx=600,bbury=800, clip=,angle=-90}
\vspace*{-0.1in}
\caption{\small The predicted  LO  differential  cross  section 
$d \sigma^{pp} / dy d^2 p_T$ for inclusive neutral pion 
and charged hadron  production at midrapidity $y=0$ in 
$p+p$ ($\bar{p}+p$) reactions  is shown for $\sqrt{s}=17, 200,$ 
and  $5500$~GeV. The ratio of neutral pions to inclusive charged 
hadrons versus  $p_T$  is given in the right panel.} 
\label{lhc-h:fig2}
\end{center} 
\end{figure}

In Fig.~\ref{lhc-h:fig2} the predicted transverse momentum 
distribution of neutral pions and inclusive charged hadrons is shown,  
corresponding to the quadratic in $\ln s$ fit to $K_{NLO}$ for 
energies typical of SPS, RHIC, and the LHC. The 
{\em significant} hardening of the spectra with $\sqrt{s}$ has 
two important consequences for $p+A$ and $A+A$ collisions: a notably 
reduced sensitivity to initial state kinematic effects 
(smaller Cronin) and larger variation of the manifested final-state 
multi-parton scattering (energy loss)  with $p_T$~\cite{Vitev:2002pf}. 
We have also investigated the effect of isospin asymmetry between 
$p+p$ and $p+\bar{p}$ reactions in $\pi^0$ and $h^+ + h^-$  
production and found it to be small. More quantitatively,  at 
$\sqrt{s}=5.5$~TeV  the fractional difference 
$|d\sigma^{\bar{p}p}-d\sigma^{pp}|/d\sigma^{pp}$ 
varies from 2.5\% at $p_T = 5$~GeV to 4.8\% at $p_T = 150$~GeV. 
This is insignificant as compared to the projected 50\% 
uncertainty that comes from the extrapolation of 
$K_{NLO}$ in LO calculations (see Fig.~\ref{lhc-h:fig1}) or the 
choice of scale in NLO calculations. A recent study  
showed {\em no} deviation from DGLAP evolution, Eq.~(\ref{mq}),   
at $Q^2=10$~GeV$^2$  down to $x=10^{-5}$ in $N+N$ 
reactions~\cite{Eskola:2002yc} The nuclear 
amplification effect $\propto A^{1/3} \simeq 10$ for a large nucleus 
is still insufficient  to enable measurements of high initial 
gluon density QCD at RHIC,  but will play an important role 
at the LHC.

\subsection{Perturbative QCD hadron composition}

The predicted hadron composition in $p+p$ ($\bar{p}+p$) reactions is plotted
in the right panel of  Fig.~\ref{lhc-h:fig2}. The proton+kaon
fraction is seen to increase systematically with $p_T$
($x_T = 2p_T/\sqrt{s}$) and is reflected in the decreasing 
$\pi^0/0.5(h^+ + h^-)$. At RHIC and LHC energies this ratio becomes 
$\sim 0.5$ at  $p_T \simeq 15$~GeV and  $p_T \simeq 75$~GeV, respectively.     
At transverse momenta $p_T \simeq 2-4$~GeV the contribution of baryons and
kaons to $h^+ + h^-$ is  $\leq 20\%$. This is significantly 
smaller compared to  data on $N+N$ reactions,
 with the discrepancy being amplified  in central $A+A$. 
Possible explanations include: enhanced  baryon production  
via topological gluon configurations (junctions)  and its interplay 
with jet quenching~\cite{Wang:xy,Gyulassy:1993hr} in 
$A+A$~\cite{Vitev:2001zn,Vitev:2002wh}, hydrodynamic 
transverse flow~\cite{Teaney:2001av},  uncertainty of the 
fragmentation functions $D_{p/c}(z_c,Q^2)$ into protons and 
antiprotons~\cite{Zhang:2002py}, and quark recombination driven 
by unorthodox (extracted) parton distributions inside 
nuclei~\cite{Hwa:2002tu}. The approaches in 
Refs.~\cite{Vitev:2001zn,Vitev:2002wh,Teaney:2001av} also address the 
centrality dependence of the baryon/meson ratios in heavy ion collisions 
at RHIC.  In~\cite{Vitev:2002wh} in has been shown 
that similar nuclear enhancement is
expected in $\Lambda, \bar{\Lambda}$ production (as compared to kaons). 
The combined low-$p_T$ baryon  enhancement and the growth of the 
non-pionic hadron fraction in the 
perturbative regime may lead to an approximately constant  pion to 
charged hadron ratio in the full measured $p_T$ region  at RHIC at 
$\sqrt{s}_{NN}=200$~GeV.  We propose that the LHC may play 
a critical role in resolving the mystery of enhanced baryon production 
in $A+A$ through the significantly larger experimentally 
accessible $p_T$ range. Effects associated  with baryon transport  and 
transverse flow are not expected to extend  beyond $p_T=10-15$~GeV and may 
result in a detectable minimum of the baryon/meson ratio versus
$p_T$  before a secondary subsequent rise. On the other hand, 
fragmentation functions (possibly enhanced at 
large $z_c$ relative to current  parameterizations) are expected 
to exhibit a much more monotonic behavior.

\section{NUCLEAR MODIFICATION FACTORS}

\begin{figure}[htb!]
\begin{center} 
\hspace*{0.3in}\epsfig{file=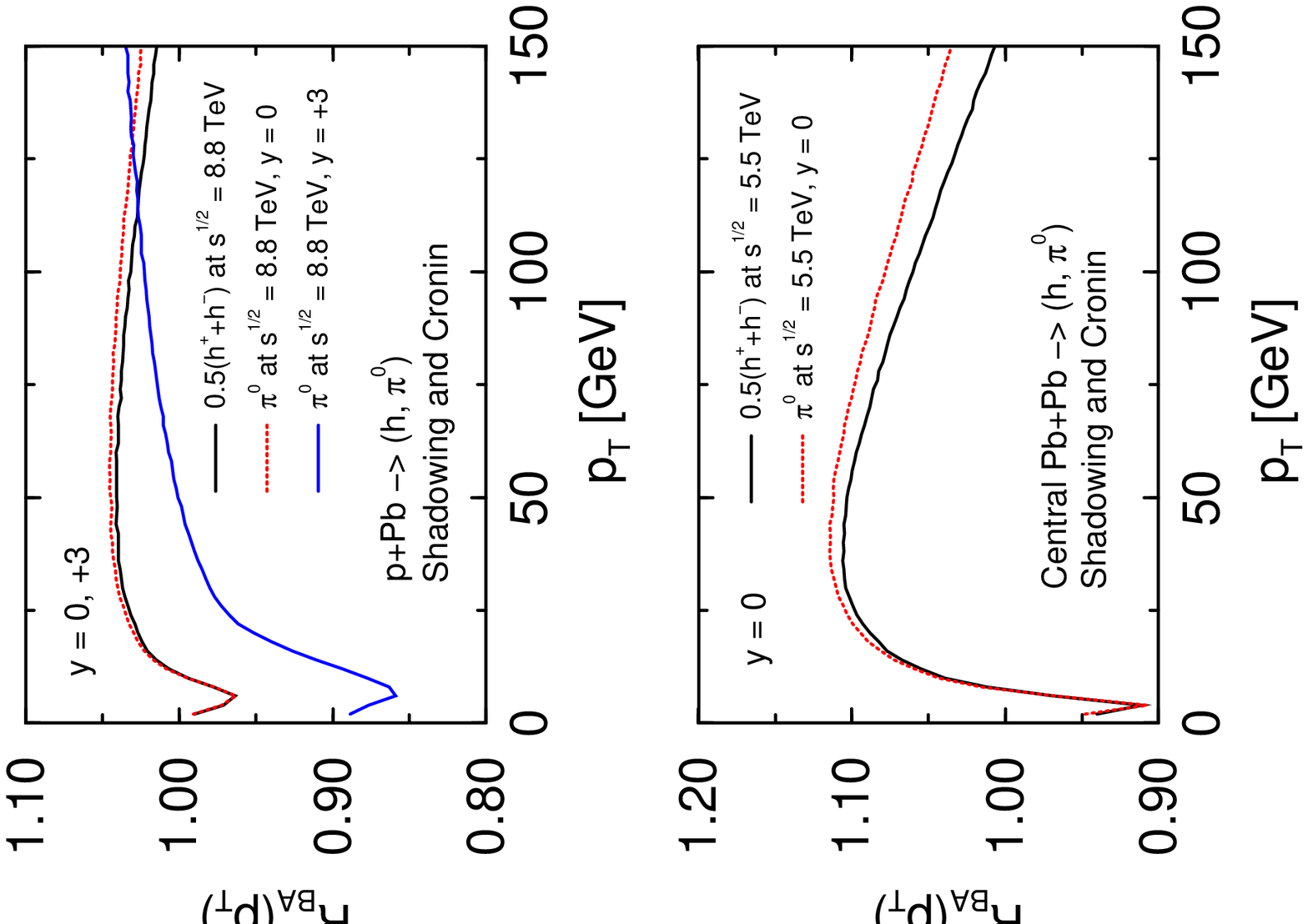,height=2.7in,width=4.9in
,angle=-90}
\hspace*{0.1in}
\epsfig{file=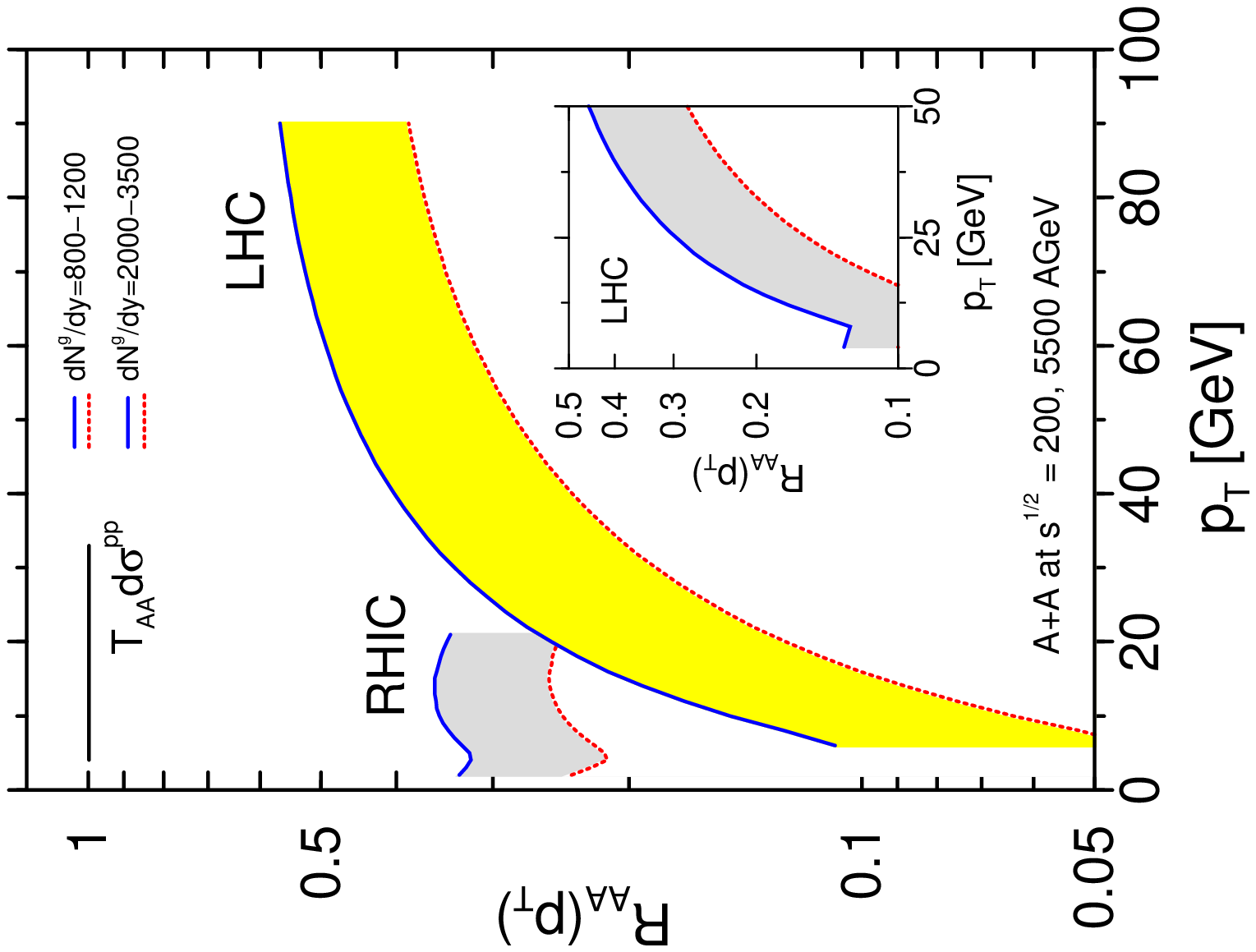,height=3.in,width=4.9in,
angle=-90}
\vspace*{0.1in}
\caption{\small The antishadowing and Cronin effects 
in $p+Pb$ and central $Pb+Pb$ without energy loss  at 
the LHC ($\sqrt{s}= 5.5$ and $8.8$~TeV)  are shown in the 
left top and bottom panels. The right panel demonstrates
the dominance of final state radiative energy loss effects 
at the LHC with  a much stronger $p_T$ dependence compared 
to RHIC. The possible restoration  of the participant scaling 
through hydrodynamic-like feedback  at $p_T\rightarrow 0$ 
is also shown~\cite{Vitev:2002pf}.  }
\label{lhc-h:fig3}
\end{center} 
\end{figure}

Dynamical  nuclear  effects  in $p+A$ and $A+A$ reactions     
are detectable through  the nuclear modification  ratio 
\begin{equation}         
R_{BA}(p_T)  = \left\{   
\begin{array}{ll}  \displaystyle 
\frac{d \sigma^{pA}}{dyd^2{p}_T} / 
\frac{ A \; d \sigma^{pp}} {dyd^2{p}_T} \;\; &  
{\rm in}  \; \;    p+A  \\[2.8ex]  \displaystyle 
\frac{d N^{AA}(b)}{dyd^2{p}_T} / 
\frac{ T_{AA}(b)\; d \sigma^{pp}}{dyd^2{p}_T}
   \;\;  &    {\rm in} \;  \; A+A  \;  \end{array} \right. ,  
\label{geomfact} 
\end{equation}          
where $A$  and $T_{AA}({b}) = 
\int  d^2{\bf r} \, T_A({\bf r})T_B({\bf r}-{\bf b})$ 
in terms  of nuclear thickness functions 
$T_A(r)=\int dz \,\rho_A({\bf r},z)$ 
are the corresponding Glauber scaling factors~\cite{Glauber:1970jm} 
of $d\sigma^{pp}$.  We note that in $R_{BA}(p_T)$ the uncertainty 
associated with the $K_{NLO}$ factors, discussed in the previous 
section, drops out.  The reference calculations that follow  
include shadowing/antishadowing/EMC-effect (here referred to as 
``shadowing''), the Cronin effect, and the non-Abelian energy 
loss of jets. The scale  dependent nuclear 
PDFs read: $f_{\alpha/A}(x,Q^2) = S_{\alpha/A} (x,Q^2) \,  ( Z/A \,  
 f_{\alpha/p}(x,Q^2) +  N/A \, f_{\alpha/n}(x,Q^2) )$,  where we 
take the isospin  effects on average and the EKS'98 
parameterization~\cite{Eskola:1998df} of the  shadowing  
functions  $S_{\alpha/A} (x,Q^2)$.  Initial   state 
multiple elastic  scatterings  have been discussed 
in~\cite{Accardi:2001ih,Qiu:2001hj,Gyulassy:2002yv}. 
From~\cite{Gyulassy:2002yv} the transverse momentum distribution 
of partons  that have undergone an average  $\chi = L/\lambda$  
incoherent interactions  in the medium can be evaluated 
exactly for any initial flux $dN^{(0)}({\bf p})$:
\begin{equation}
dN({\bf p}) = \sum_{n=0}^\infty e^{-\chi} \frac{\chi^n}{n!}  
\int \prod_{i=1}^n d^2 {\bf q}_i \frac{1}{\sigma_{el}} 
 \frac{d \sigma_{el} }{d^2 {\bf q}_i}  \;  
dN^{(0)}({\bf p} -{\bf q}_1 - \cdots - {\bf q}_n)  \;. 
\label{glauber}
\end{equation} 
Numerical estimates of~(\ref{glauber}) 
show that for thin media with a few  semi-hard  
scatterings  the induced  transverse  momentum   broadening 
exhibits a weak logarithmic enhancement  with $p_T$  and  is
proportional to $L \propto A^{1/3}$. The transverse 
momentum transfer per unit length in cold nuclear matter 
is found to be  
$\mu^2 / \lambda \simeq 0.05$~GeV$^2$/fm~\cite{Vitev:2002pf}
from comparison to low energy $p+A$ 
data~\cite{Cronin:zm,Straub:xd,Antreasyan:cw}.  The left top and 
bottom panels of Fig.~\ref{lhc-h:fig3} show the predicted 
Cronin+shadowing effect in $p+Pb$ collisions at $\sqrt{s}=8.8$~TeV
and  central $Pb+Pb$ at $\sqrt{s}=5.5$~TeV without final 
state medium induced energy loss.  The 4\% (10\%) enhancement 
of $R_{BA}$ at $p_T \simeq 40$~GeV comes from antishadowing and in 
not related no multiple initial state scattering. 
The observed difference  between $\pi^0$ and $0.5 (h^+ + h^-)$ reflects 
the different  $S_\alpha(x,Q^2)$ for quarks and gluons. 
Cronin effect at  the LHC results  in  slowing 
down of the decrease of $R_{BA}$ at  small $x$ as seen 
in  the  $p_T \rightarrow 0$  limit. In contrast, at  RHIC  one  
finds  $\simeq 30\%$  enhancement in $d+Au$ reactions  
at $\sqrt{s}=200$~GeV  and  $\simeq 60\%$  effect in central 
$Au+Au $  relative  to  the  {\em binary collision}  scaled  $p+p$  
result. At CERN-SPS energies of $\sqrt{s}= 17$~GeV  the  results 
are most striking, with values reaching 250\% in $d+Au$ 
and 400\%  in central $Au+Au$ at $ p_T \simeq 4$~GeV.  
For a summary of results on midrapidity Cronin effect at the LHC 
see~\cite{A}.

The manifestation of multiple initial state scattering  
and nuclear shadowing at forward and backward rapidities 
$y=\pm 3$ in $p+Pb$ at the LHC (for CMS $\eta \leq 2.5$)   
and $d+Au$  at RHIC (for BRAHMS $ \eta \leq 3$ ) has also 
been studied in the framework of a fixed (or slowly varying)  
initial parton interaction strength. At  LHC  energies at  
$y=+3$ (in the direction of the proton beam) the effect of the 
sequential projectile interactions is again small (due to the much 
flatter rapidity and transverse momentum distributions) and 
is overwhelmed  by  shadowing,  which is found  to  be a factor 
of  2-3  times larger than the $y=0$ result 
at small $p_T \sim$ few GeV and vanishes ($R_{BA}=1$) at 
$p_T \simeq 50$~GeV. As previously emphasized, initial state gluon 
showering can significantly change the low-$p_T$ behavior of 
the hadronic spectra at the LHC beyond  the  current  shadowing 
parameterization. At RHIC in $d+Au$ reactions  at $\sqrt{s} = 200$~GeV 
the nuclear modification ratio is qualitatively different. 
While near nucleus beam  (backward $y=-3$)  rapidity 
$R_{BA} \simeq 0.9 - 1$ at forward rapidities $y=+3$ the nuclear 
modification factor exhibits a  much more dramatic $p_T$ dependence. 
At  small transverse momenta  $p_T \sim 1$~GeV  hadron 
production is suppressed  relative to the
binary collision scaled $p+p$ result, $R_{BA} \leq 0.8$. The maximum 
Cronin enhancement $R_{BA}^{\max} \simeq 1.3$ (30\%) is essentially the
same as at midrapidity~\cite{Vitev:2002pf} but slightly  shifted 
to larger $p_T$. We emphasize  that {\em both} the suppression 
and enhancement regions are an integral 
part of the Cronin effect~\cite{Cronin:zm,Straub:xd,Antreasyan:cw} 
that is understood in terms of probability  conservation and momentum 
redistribution resulting from  multiple  initial state 
scattering~\cite{Vitev:2002pf,Zhang:2001ce,Glauber:1970jm,Accardi:2001ih,Qiu:2001hj,Gyulassy:2002yv,A}.  
At forward (in the direction of the deuteron beam) rapidities a calculation as 
in~\cite{Vitev:2002pf} demonstrates a {\em  broader} Cronin enhancement 
region with $R_{BA} \simeq 25 \%$  at 
$p_T=5$~GeV. This is understood in terms of the significantly 
steeper fall-off of the hadron spectra away from midrapidity that 
enhances the effect of the otherwise similar transverse momentum kicks.  
While the discussed moderate $p_T$ interval lies at 
the very  edge of BRAHMS acceptance (at $y=+3$) the same qualitative  
picture holds at $y=+2$.

\vskip 0.5cm

The full solution for the medium induced gluon radiation 
off jets produced in a hard collisions inside  the  nuclear  
medium  of length $L$ and computed {\em to all orders} 
in the correlations  between the  multiple  scattering  centers  
via the GLV  reaction operator approach~\cite{Gyulassy:2000er}   
can be written as ($x=k^+/p^+ \approx \omega/ E$)
\begin{eqnarray}
 \sum\limits_{n=1}^\infty  x\frac{dN^{(n)}}{dx\, d^2 {\bf k}}   
& = & \frac{C_R \alpha_s}{\pi^2} \sum\limits_{n=1}^{\infty} 
 \prod_{i=1}^n \int\limits_0^{L-\sum_{a=1}^{i-1} \Delta z_a } 
 \frac{d \Delta z_i }{\lambda_g(i)} 
 \int \prod_{i=1}^n         \left(d^2{\bf q}_{i} \, 
\left[ |\bar{v}_i({\bf q}_{i})|^2 - \delta^2({\bf q}_{i}) \right]\right) 
\,  \nonumber \\[1.ex] 
&\;& \times 
\left( -2\,{\bf C}_{(1, \cdots ,n)} \cdot 
\sum_{m=1}^n {\bf B}_{(m+1, \cdots ,n)(m, \cdots, n)} 
\right . \nonumber \\[1.ex] &\;&  \left. \times \;
\left[ \cos \left (
\, \sum_{k=2}^m \omega_{(k,\cdots,n)} \Delta z_k \right)
-   \cos \left (\, \sum_{k=1}^m \omega_{(k,\cdots,n)} \Delta z_k \right)
\right]\; \right) \;, \quad \qquad  
\label{difdistro} 
\end{eqnarray}
where $\sum_2^1 \equiv 0$ is understood. In (\ref{difdistro})  
${\bf C}_{(m, \cdots ,n)} =  \frac{1}{2} \nabla_{{\bf k}} 
\ln ({\bf k} - {\bf q}_m - \cdots  - {\bf q}_n )^2$,   
${\bf B}_{(m+1, \cdots ,n)(m, \cdots, n)} = 
{\bf C}_{(m+1, \cdots ,n)} - {\bf C}_{(m, \cdots ,n)}$ are the 
color current propagators, $\omega_{(m,\cdots,n)}^{-1}  = 
2 x E / |C_{(m,\cdots,n)}^2 | $ are formation times, and 
$ \Delta z_k = z_k - z_{k-1} $ are the separations of subsequent 
scattering centers. The momentum transfers ${\bf q}_i$ are distributed 
according to a normalized  elastic scattering cross section  
$|\bar{v}_i({\bf q}_{i})|^2 = 
\sigma_{el}^{-1}d \sigma_{el} / d^2 {\bf q}_i  $  
and  the  radiative spectrum can   be evaluated from 
(\ref{difdistro}) for any initial  nuclear geometry 
with an arbitrary  subsequent  dynamical evolution 
of the matter density. At large jet energies  the lowest order 
correlation between the jet production point one of the 
scatterings that follow has been shown to dominate and 
lead to a quadratic  mean energy loss  
dependence on the size of the plasma, $\Delta E \propto L^2$  
for {\em static} media~\cite{Gyulassy:2000fs}.  To improve 
the numerical accuracy for small parton energies  we 
include corrections to third order in opacity~\cite{Vitev:2002pf}. 
The dynamical expansion of the bulk soft matter is assumed to be 
of Bjorken type. For a summary of results from recent non-Abelian 
energy loss calculations see~\cite{BLVW}.

In the Poisson approximation of independent gluon 
emission~\cite{Baier:2001yt,Gyulassy:2001nm,Wang:2002ri,Salgado:2002cd}   
the probability distribution $P(\epsilon,E)$ of the fractional 
energy loss  $\epsilon=\sum_i \omega_i/E$ can be obtained iteratively 
from the single inclusive gluon radiation  spectrum 
$dN(x,E)/dx$~\cite{Gyulassy:2001nm}. If a fast parton looses 
 $\epsilon \, E$ of its initial energy prior 
to hadronization its momentum fraction $z_c$ is modified to  
$z_c^* = p_h/p_c(1-\epsilon) = z_c/(1-\epsilon)$. The observable 
suppressed  hadron  differential cross section can be 
computed  from  Eq.~(\ref{hcrossec}) with the substitution 
\begin{equation} 
D_{h/c}(z_c,Q^2) \longrightarrow  \int d \epsilon  \, P(\epsilon,p_c) 
\frac{z_c^*}{z_c} D_{h/c}(z^*_c,Q^2)    \; . 
\label{fragmod}
\end{equation}   
The nuclear  modification  factor $R_{AA}(p_T)$ at the LHC is 
shown on the right panel of Fig.~\ref{lhc-h:fig3} and is completely 
dominated by final state interactions (see left panel). It shows a 
{\em significantly stronger}  $p_T$  dependence as compared to RHIC,
where jet quenching was predicted to be {\em approximately 
constant} over the full measured 
moderate- to high-transverse momentum range~\cite{Vitev:2002pf},   
the result of an interplay of shadowing, Cronin effect, and  
radiative energy loss.  The  variation of  $R_{AA}$ at the LHC 
is a  factor of 5: from  10-20 fold suppression 
at $p_T = 10$~GeV to only a factor  2-3 suppression 
at $p_T = 100$~GeV. The reason for such a prominent variation   
is the hardening of the particle transverse momentum spectra and the  
insufficient  balancing action  of multiple initial state scatterings. 
In fact the prediction from  Fig.~\ref{lhc-h:fig3}  is  that 
the suppression  in central $Pb+Pb$ at $\sqrt{s}_{NN}=5.5$~TeV 
at $p_T \simeq 40$~GeV is comparable to the factor of 4-5 
suppression currently observed at RHIC.

The extrapolation of the LHC quenching calculations to small 
$p_T \rightarrow 0$ results into suppression below participant scaling. 
More careful examination of the mean energy loss of partons, 
in particular for  gluons radiating in  nuclear matter at
LHC densities, reveals sizable regions of phase  space 
with $\Delta E \geq E$. This indicates complete absorption of 
jets in nuclear matter. There is experimental evidence that this 
regime of extreme {\em final state} densities may have been 
achieved at RHIC~\cite{Mioduszewski:2002wt,Jacobs:2002pz,Kunde:2002pb}.
In this case Eq.~(\ref{fragmod}) has to be corrected to 
include the feedback of the radiated gluons into the system. 
This hydrodynamic-like feedback  is expected to recover
the $N_{part}$ scaling  in the soft $p_T$  region~\cite{Vitev:2002pf} - 
also illustrated  on the right panel of Fig.~\ref{lhc-h:fig3}. 
The effective initial gluon density derived from the rapidity  
densities used in Fig.~\ref{lhc-h:fig3} are $\rho_g(RHIC) = 
30-50$/fm$^3$ and $\rho_g(LHC) = 130-275$/fm$^3$. These are 
one to two  orders of magnitude larger than  the density 
of cold nuclear matter and are suggestive of a deconfined 
QCD state  -  the quark-gluon plasma.  Interestingly, 
a recent study of  non-equilibrium  parton transport 
in central $Au+Au$ and $Pb+Pb$ at $\sqrt{s}_{NN}= 200$~GeV and 
$\sqrt{s}=5.5$~TeV  has found  initial  parton  densities  
corresponding to the lower bound of the intervals quoted above.

\section{CONCLUSIONS}

In summary, a lowest order pQCD analysis of single inclusive 
hadron production  has  been  performed, revealing a systematic 
decrease with $\sqrt{s}$ of the contribution of the 
next-to-leading corrections to the differential cross sections.
The predicted $d\sigma^h / dy d^2 p_T$ exhibits significant 
hardening with transverse momentum and an increased fractional
contribution of kaons and protons at high $p_T$, the latter also being 
true at RHIC energies. In central $A+A$ reactions  the nuclear 
modification factor $R_{AA}(p_T)$ at the LHC is shown to be  
completely dominated by final state multi-parton 
interactions~\cite{Vitev:2002pf}. For comparison, at RHIC 
Cronin effect and nuclear shadowing also play an important role, 
leading to an approximately constant suppression ratio. At the SPS  
initial state multiple elastic scatterings dominate, resulting 
in a net enhancement of hadron production.  At forward ($y=+3$) 
rapidities in $d+Au$ at RHIC the Cronin enhancement region is predicted 
to be broader in comparison to the $y=0$ case. In contrast in $p+Pb$ at 
the LHC nuclear shadowing dominates but in order to detect a sizable  
reduction relative to the binary collision scaled $p+p$ cross section 
measurements at close to proton rapidity ($y_{\max}=9.2$ for 
$\sqrt{s}=8.8$~TeV) are needed. 
    
The predicted decreasing $R_{AA}$ with  $p_T$ at the LHC, if confirmed, 
may have important experimental consequences. Comparative 
large-$E_T$ measurements of the difference in the 
{\em full structure} of the jet cone in $p+p$ and $A+A$ reactions 
may  prove difficult for weak signals and large
backgrounds. We emphasize that one of the easiest 
and most unambiguous approaches 
for detecting  the non-Abelian jet energy loss  and  performing 
jet-tomographic analysis of the properties of the hot and dense 
matter created  in ultra-relativistic heavy ion reactions is 
through  the suppression pattern of leading hadrons. Therefore these 
measurements should enter as an important part of the experimental
programs at the LHC.         
 
\vspace{.4cm}

\noindent {\bf Acknowledgments}

\noindent  Many helpful discussions with  M.~Gyulassy, K.~Eskola, 
V.~Kolhinen,  J.~W.~Qiu, and J.~Vary are  gratefully acknowledged. 
This work is supported  by the United States Department of Energy 
under  Grant No. DE-FG02-87ER40371.


\begin{thebibliography}{99} 

\bibitem{Collins:1974ky}
J.~C.~Collins and M.~J.~Perry,
Phys.\ Rev.\ Lett.\  {\bf 34} (1975) 1353.


\bibitem{Mueller:wy}
A.~H.~Mueller and J.~w.~Qiu,
Nucl.\ Phys.\ B {\bf 268} (1986) 427.


\bibitem{Dumitru:2002qt}
A.~Dumitru and J.~Jalilian-Marian,
Phys.\ Rev.\ Lett.\  {\bf 89} (2002) 022301
[arXiv:hep-ph/0204028].




\bibitem{Vitev:2002pf}
I.~Vitev and M.~Gyulassy,
Phys.\ Rev.\ Lett.\ {\bf 89} (2002) 252301 [arXiv:hep-ph/0209161].




\bibitem{Accardi:2002vt}
A.~Accardi, N.~Armesto and I.~P.~Lokhtin,
{\em these proceedings}, arXiv:hep-ph/0211314.

  
\bibitem{nloh} I.~Sarcevic, {\em these proceedings}; 
P. Levai and  G.G.~Barnafoldi, {\em these proceedings}.


\bibitem{Owens:1986mp}
J.~F.~Owens,
Rev.\ Mod.\ Phys.\  {\bf 59} (1987) 465.




\bibitem{Cronin:zm} 
J.~W.~Cronin, H.~J.~Frisch, M.~J.~Shochet, 
J.~P.~Boymond, R.~Mermod, P.~A.~Piroue and R.~L.~Sumner,
in {\it C74-07-01.17}
Phys.\ Rev.\ D {\bf 11} (1975) 3105.



\bibitem{Straub:xd}
P.~B.~Straub {\it et al.},
Phys.\ Rev.\ Lett.\  {\bf 68} (1992) 452.



\bibitem{Antreasyan:cw}
D.~Antreasyan, J.~W.~Cronin, H.~J.~Frisch, M.~J.~Shochet, 
L.~Kluberg, P.~A.~Piroue and R.~L.~Sumner,
Phys.\ Rev.\ D {\bf 19} (1979) 764.




\bibitem{Alper:nv}
B.~Alper {\it et al.}  [British-Scandinavian ISR Collaboration],
Phys.\ Lett.\ B {\bf 44} (1973) 521.


\bibitem{Albajar:1989an}
C.~Albajar {\it et al.}  [UA1 Collaboration],
Nucl.\ Phys.\ B {\bf 335} (1990) 261.


\bibitem{Banner:1984wh}
M.~Banner {\it et al.}  [UA2 Collaboration],
Z.\ Phys.\ C {\bf 27} (1985) 329.


\bibitem{Bocquet:1995jr}
G.~Bocquet {\it et al.},
Phys.\ Lett.\ B {\bf 366} (1996) 434.


\bibitem{Abe:yu}
F.~Abe {\it et al.}  [CDF Collaboration],
Phys.\ Rev.\ Lett.\  {\bf 61} (1988) 1819.









\bibitem{Zhang:2001ce}
Y.~Zhang, G.~Fai, G.~Papp, G.~G.~Barnafoldi and P.~Levai,
Phys.\ Rev.\ C {\bf 65} (2002) 034903
[arXiv:hep-ph/0109233].


\bibitem{Eskola:2002kv}
K.~J.~Eskola and H.~Honkanen,
Nucl.\ Phys.\ A {\bf 713} (2003) 167
[arXiv:hep-ph/0205048].








\bibitem{Gluck:1998xa}
M.~Gluck, E.~Reya and A.~Vogt,
Eur.\ Phys.\ J.\ C {\bf 5} (1998) 461
[arXiv:hep-ph/9806404].


\bibitem{Binnewies:1994ju}
J.~Binnewies, B.~A.~Kniehl and G.~Kramer,
Z.\ Phys.\ C {\bf 65} (1995) 471
[arXiv:hep-ph/9407347].



\bibitem{Wang:1998bh}
X.~N.~Wang,
Phys.\ Rev.\ C {\bf 58} (1998) 2321
[arXiv:hep-ph/9804357].



\bibitem{Sjostrand:1993yb}
T.~Sjostrand,
Comput.\ Phys.\ Commun.\  {\bf 82} (1994) 74.




\bibitem{Eskola:2002yc}
K.~J.~Eskola, H.~Honkanen, V.~J.~Kolhinen, J.~Qiu and C.~A.~Salgado,
arXiv:hep-ph/0211239.





\bibitem{Wang:xy}
X.~N.~Wang and M.~Gyulassy,
Phys.\ Rev.\ Lett.\  {\bf 68} (1992) 1480.


\bibitem{Gyulassy:1993hr}
M.~Gyulassy and X.~n.~Wang,
Nucl.\ Phys.\ B {\bf 420} (1994) 583
[arXiv:nucl-th/9306003].










\bibitem{Vitev:2001zn}
I.~Vitev and M.~Gyulassy,
Phys.\ Rev.\ C {\bf 65} (2002) 041902
[arXiv:nucl-th/0104066].


\bibitem{Vitev:2002wh}
I.~Vitev and M.~Gyulassy,
arXiv:hep-ph/0208108.



\bibitem{Teaney:2001av}
D.~Teaney, J.~Lauret and E.~V.~Shuryak,
arXiv:nucl-th/0110037.


\bibitem{Zhang:2002py}
X.~f.~Zhang, G.~Fai and P.~Levai,
arXiv:hep-ph/0205008.


\bibitem{Hwa:2002tu}
R.~C.~Hwa and C.~B.~Yang,
arXiv:nucl-th/0211010.








\bibitem{Glauber:1970jm}
R.~J.~Glauber and G.~Matthiae,
Nucl.\ Phys.\ B {\bf 21} (1970) 135.







\bibitem{Eskola:1998df}
K.~J.~Eskola, V.~J.~Kolhinen and C.~A.~Salgado,
Eur.\ Phys.\ J.\ C {\bf 9} (1999) 61
[arXiv:hep-ph/9807297].




\bibitem{Accardi:2001ih}
A.~Accardi and D.~Treleani,
Phys.\ Rev.\ D {\bf 64} (2001) 116004
[arXiv:hep-ph/0106306].


\bibitem{Qiu:2001hj}
J.~w.~Qiu and G.~Sterman,
arXiv:hep-ph/0111002.



\bibitem{Gyulassy:2002yv}
M.~Gyulassy, P.~Levai and I.~Vitev,
Phys.\ Rev.\ D {\bf 66} (2002) 014005
[arXiv:nucl-th/0201078].


\bibitem{A}
For a summary of recent results on the Cronin effect see 
A.~Accardi, {\em these proceedings}.


\bibitem{Gyulassy:2000er}
M.~Gyulassy, P.~Levai and I.~Vitev,
Nucl.\ Phys.\ B {\bf 594} (2001) 371
[arXiv:nucl-th/0006010].




\bibitem{Gyulassy:2000fs}
M.~Gyulassy, P.~Levai and I.~Vitev,
Phys.\ Rev.\ Lett.\  {\bf 85} (2000) 5535
[arXiv:nucl-th/0005032].



\bibitem{BLVW}
For  a summary of recent results on non-Abelian energy loss see
R.~Baier, P.~Levai, I.~Vitev, U.~A.~Wiedemann, {\em these 
proceedings}.  





\bibitem{Baier:2001yt}
R.~Baier, Y.~L.~Dokshitzer, A.~H.~Mueller and D.~Schiff,
JHEP {\bf 0109} (2001) 033
[arXiv:hep-ph/0106347].


\bibitem{Gyulassy:2001nm}
M.~Gyulassy, P.~Levai and I.~Vitev,
Phys.\ Lett.\ B {\bf 538} (2002) 282
[arXiv:nucl-th/0112071].


\bibitem{Wang:2002ri}
E.~Wang and X.~N.~Wang,
Phys.\ Rev.\ Lett.\  {\bf 89} (2002) 162301
[arXiv:hep-ph/0202105].


\bibitem{Salgado:2002cd}
C.~A.~Salgado and U.~A.~Wiedemann,
Phys.\ Rev.\ Lett.\  {\bf 89} (2002) 092303
[arXiv:hep-ph/0204221].




\bibitem{Mioduszewski:2002wt}
S.~Mioduszewski  [PHENIX Collaboration],
arXiv:nucl-ex/0210021.




\bibitem{Jacobs:2002pz}
P.~Jacobs,
arXiv:hep-ex/0211031.


\bibitem{Kunde:2002pb}
G.~J.~Kunde,
arXiv:nucl-ex/0211018.




\bibitem{Cooper:2002td}
F.~Cooper, E.~Mottola and G.~C.~Nayak,
arXiv:hep-ph/0210391.



\end{thebibliography}
\end{document}